\documentclass[A4,12pt]{article}

\usepackage{cite}

\newcommand{\tz}[0]{{{}^3\!\!/\!{}_2 }}

\title{On massive spin-$\tz$ 
in the Fradkin-Vasiliev formalism}

\author{M.V. Khabarov\thanks{maksim.khabarov@ihep.ru},
Yu.M. Zinoviev\thanks{Yurii.Zinoviev@ihep.ru}
\\[0.5cm]
\it{\small Institute for High Energy Physics of National
Research Center "Kurchatov Institute"} \\
\it{\small Protvino, Moscow Region, 142281, Russia}}

\date{}

\begin{document}

\maketitle

\begin{abstract}
One of the possible approaches to the construction of massive higher
spin interactions is to use their gauge invariant description based on
the introduction of the appropriate set of Stueckelberg fields.
Recently, the general properties of such approach were investigated in
\cite{BDGT18}. The main findings of this work can be formulated in two
statements. At first, there always exist enough field redefinitions to
bring the vertex into abelian form where there are some corrections to
the gauge transformations but the gauge algebra is undeformed. At
second, with the further (as a rule higher derivative) field
redefinitions one can bring the vertex into trivially gauge invariant
form expressed in terms of the gauge invariant objects of the free
theory.

Our aim in this work is to show (using a simple example) how these
general properties are realised in the so-called Fradkin-Vasiliev
formalism and to see the effects (if any) that the presence of
massless field, and hence of some unbroken gauge symmetries, can
produce. As such example we take the gravitational interaction for
massive spin-$\tz$ field so we complete the investigation started in
\cite{BSZ14} relaxing all restrictions on the number of derivatives
and allowed field redefinitions. We show that in spite of the
presence of massless spin-2 field, the first statement is still valid,
while there exist two abelian vertices which are not equivalent 
on-shell to the trivially gauge invariant ones. Moreover, it is one of
this abelian vertices that reproduce the minimal interaction for
massive spin-$\tz$.
\end{abstract}

\thispagestyle{empty}
\newpage
\setcounter{page}{1}

\section{Introduction}

Gauge invariance plays a crucial role in the construction of the
consistent interaction vertices for massless higher spin $s \ge 1$
fields. At the same time, massive fields in their minimal covariant
formulation do not have any gauge symmetry so that one has to look for
other approaches. But the gauge invariant description for massive
fields, where gauge invariance is achieved with the introduction of
the appropriate set of Stueckelberg fields, also exists both for the
metric-like \cite{Zin01,Met06} as well as frame-like
\cite{Zin08b,PV10,KhZ19} formalism. Thus we can try to follow the
same line as for the massless fields though the procedure appears to
be quit different in many aspects. Recently, the general properties of
such approach were investigated \cite{BDGT18} using metric-like
formalism with some concrete examples as illustrations. Let us briefly
formulate the main finding of this work in the language we adopt in
this paper. First statement (or first part of the theorem if one
likes) is that there always exist enough field redefinitions to bring
the vertex into the abelian form\footnote{We call the vertex to be
trivially gauge invariant if it does not change initial gauge
transformations; abelian if it changes the gauge transformations but
not the algebra; non-abelian if it changes both the gauge
transformations and the algebra.}. The second statement (part) is that
 with the further (as a rule higher derivative) field redefinitions
one can bring the vertex into trivially gauge invariant form expressed
in terms of gauge invariant objects of the free theory. Let us stress
here that all these three forms (non-abelian, abelian and trivially
gauge invariant) are equivalent on-shell.

Our aim in this work is twofold. First of all, using the most simple
(after Yang-Mills theory) example to show how these general
properties are realised in the so-called Fradkin-Vasiliev approach
\cite{FV87,FV87a,Vas11} based on the frame-like formalism. The second
point is to see the effects (if any) that the presence of massless
field, and hence some unbroken gauge symmetries, can produce. As such
example we take the gravitational interaction for massive spin-$\tz$
field. This case has been already considered in \cite{BSZ14} with the
implicit restrictions on the number of derivatives and allowed field
redefinitions\footnote{Note that there were other applications of such
formalism where (explicitly or implicitly) restrictions on the number
of derivatives and allowed field redefinitions were imposed
\cite{Zin10a,Zin11,Zin14}.}. Here we complete this investigation
relaxing any such constraints. 

Let us recall the main points of the Fradkin-Vasiliev formalism and
compare its application to massless and massive cases. Massless fields
in the frame-like formalism is described by a set of one-forms $\Phi$
(physical, auxiliary and extra ones) each one playing the role of the
gauge fields with its own gauge transformations
$$
\delta \Phi \sim D \xi + \dots
$$
For each field a gauge invariant two-form (curvature) can be
constructed
$$
{\cal R} \sim D \Phi + \dots
$$
On-shell most of these curvatures vanish so that all auxiliary and
extra fields are equivalent to the higher derivatives of the physical
one. At last, in $AdS$ space with the non-zero cosmological constant
the free Lagrangian can be written in an explicitly gauge invariant
form
$$
{\cal L}_0 \sim \sum {\cal R} {\cal R} 
$$
To non-linearly deform the free theory one consider the most general
quadratic deformation for all the curvatures
$$
{\cal R} \Rightarrow \hat{\cal R} = {\cal R} + \Delta {\cal R}, \qquad
\Delta {\cal R} \sim \Phi \Phi
$$
Naturally, these deformed curvatures ceased to be gauge invariant and
the main requirement on this step is that they must transform
covariantly under all gauge transformations, i.e.
$$
\delta \hat{\cal R} \sim {\cal R} \xi
$$
This allows one directly
read out the corrections to the gauge transformations
$$
\delta \Phi \sim \Phi \xi
$$
and severely restrict all parameters of the deformations. Let us
stress
that having only one-forms we cannot introduce any field
redefinitions without spoiling the very nature of the frame-like
formalism (e.g. mixing the world and local indices using the inverse
frame. Thus the procedure appears to be straightforward and
unambiguous
(and especially simple in $d=4$, see e.g. \cite{KhZ20a,KhZ20b}). At
the second step one replace the initial curvatures in the free
Lagrangian with the deformed ones and requires the Lagrangian to be
invariant. As has been shown by Vasiliev \cite{Vas11}, to construct
the full set of cubic vertices one also have to add all possible
abelian terms of the form ${\cal R}{\cal R}\Phi$ to this deformed
Lagrangian.

Now let us turn to the massive case. Frame-like gauge invariant
description for the massive fields \cite{Zin08b,PV10,KhZ19} also
requires a set of one-forms $\Phi$ as well as a similar set (there is
a one-to-one correspondence) of zero-forms $W$ playing the role of
Stueckelberg fields:
$$
\delta \Phi \sim D \xi + \dots, \qquad \delta W \sim m \xi
$$
Each of them has its own gauge invariant object (two-form or one-form
witch we collectively call curvatures):
$$
{\cal R} \sim D \Phi + \dots, \qquad 
{\cal C} \sim D W - m \Phi + \dots
$$
As in the massless case the free Lagrangian can be written in the
explicitly gauge invariant form
$$
{\cal L}_0 \sim \sum [ {\cal R} {\cal R} + {\cal R} {\cal C} +
{\cal C} {\cal C}]
$$
Note however that in the massive case this is possible not only in
$AdS$ space but also in Minkowski space and even in $dS$ space (inside
the unitary allowed region). 

Now let us turn to the interaction. The first step is again the most
general quadratic deformation for all curvatures:
\begin{eqnarray*}
\Delta {\cal R} &\sim& \Phi \Phi + \Phi W + W W \\
\Delta {\cal C} &\sim& \Phi W + W W 
\end{eqnarray*}
with the corresponding corrections to the gauge transformations:
$$
\delta \Phi \sim \Phi \xi + W \xi, \quad \delta W \sim W \xi
$$
and with the requirement that the deformed curvatures transform
covariantly. Due to the presence of zero-forms one can construct a lot
of possible field redefinitions
$$
\Phi \Rightarrow \Phi + \Phi W + W W, \qquad 
W \Rightarrow W + W W
$$
so that the procedure drastically differs form that in the massless
case.

The second step is the same as before: one has to replace the initial
curvatures in the free Lagrangian with the deformed ones
$$
{\cal L}_0 \Rightarrow \hat{\cal L} \sim \sum [\hat{\cal R}
\hat{\cal R} + \hat{\cal R} \hat{\cal C} + \hat{\cal C}
\hat{\cal C} ]
$$
add all possible abelian terms and require the total Lagrangian to be
invariant.

The paper is organized as follows. In section 2 we provide all
necessary information on the description of massless spin-2 and
massive spin-$\tz$ in $AdS_4$ background. In section 3 and 4 we
consider the most general quadratic deformations for spin-$\tz$ and
spin-2 curvatures correspondingly. In both cases we show that there
exist enough field redefinitions  to reduce the vertex to the abelian
form so that the first part of the theorem in \cite{BDGT18} works even
with the presence of massless spin-2. In section 5 we consider the
most general abelian vertex and show that two of them are independent
in a sense that they are not on-shell equivalent to the trivially
gauge invariant ones contrary to the purely massive case. Moreover, it
is one of this vertices that gives the minimal gravitational
interactions for massive spin-$\tz$. At last, in section 6 using the
field redefinitions we bring the vertex back to its non-abelian
incarnation.

\section{Kinematics} 

In this section we briefly recall all necessary information on the
description of massless spin-2 and massive spin-$\tz$ fields living in
$AdS_4$ background.

\subsection{Massless spin 2}

We work in the frame-like multispinor formalism\footnote{We use the
same notations and conventions as in our previous works 
\cite{KhZ19,KhZ20a,KhZ20b}.} where the massless spin-2 is described by
the dynamical frame one-form $h^{\alpha\dot\alpha}$ and dynamical
Lorentz connection one-forms  $\omega^{\alpha(2)}$, 
$\omega^{\dot\alpha(2)}$. The Lagrangian (four-form in our formalism)
for the free field in $AdS_4$ looks like:
\begin{equation}
\frac{1}{i} {\cal L}_0 = \omega^{\alpha\beta} E_\beta{}^\gamma
\omega_{\alpha\gamma} + \omega^{\alpha\beta} e_\beta{}^{\dot\alpha}
D h_{\alpha\dot\alpha} + 2\lambda^2 h^{\alpha\dot\alpha} 
E_\alpha{}^\beta h_{\beta\dot\alpha} + h.c.
\end{equation}
Here $e^{\alpha\dot\alpha}$ is the background frame and $D$ is the
Lorentz
covariant derivative of $AdS_4$ defined so that
$$
D e^{\alpha\dot\alpha} = 0, \qquad D D \zeta^\alpha = - 2\lambda^2
E^\alpha{}_\beta \zeta^\beta
$$
and the two-forms $E^{\alpha(2)}$, $E^{\dot\alpha(2)}$ are defined as
follows
$$
e^{\alpha\dot\alpha} e^{\beta\dot\beta} = \epsilon^{\alpha\beta}
E^{\dot\alpha\dot\beta} + \epsilon^{\dot\alpha\dot\beta}
E^{\alpha\beta}
$$
This Lagrangian is invariant under the following gauge
transformations:
\begin{eqnarray}
\delta \omega^{\alpha(2)} &=& D \eta^{\alpha(2)} 
+ \lambda^2 e^\alpha{}_{\dot\beta} \xi^{\alpha\dot\beta} \nonumber \\
\delta \omega^{\dot\alpha(2)} &=& D \eta^{\dot\alpha(2)}
+ \lambda^2 e_\beta{}^{\dot\alpha} \xi^{\beta\dot\alpha} \\
\delta h^{\alpha\dot\alpha} &=& D \xi^{\alpha\dot\alpha} 
+ e_\beta{}^{\dot\alpha} \eta^{\alpha\beta}
+ e^\alpha{}_{\dot\beta} \eta^{\dot\alpha\dot\beta} \nonumber
\end{eqnarray}
There are two gauge invariant two-forms (curvature and
torsion)\footnote{In what follows all such gauge invariant two- and
one-forms we collectively call curvatures.}:
\begin{eqnarray}
R^{\alpha(2)} &=& D \omega^{\alpha(2)} 
+ \lambda^2 e^\alpha{}_{\dot\alpha} h^{\alpha\dot\alpha} \nonumber \\
R^{\dot\alpha(2)} &=& D \omega^{\dot\alpha(2)}
+ \lambda^2 e_\beta{}^{\dot\alpha} h^{\beta\dot\alpha} \\
T^{\alpha\dot\alpha} &=& D h^{\alpha\dot\alpha}
+ e_\beta{}^{\dot\alpha} \omega^{\alpha\beta}
+ e^\alpha{}_{\dot\beta} \omega^{\dot\alpha\dot\beta} \nonumber
\end{eqnarray}
 They satisfy the following differential identities:
\begin{eqnarray}
D R^{\alpha(2)} &=& - \lambda^2 e^\alpha{}_{\dot\alpha}
T^{\alpha\dot\alpha}, \nonumber \\
D R^{\dot\alpha(2)} &=& - \lambda^2 e_\beta{}^{\dot\alpha}
T^{\beta\dot\alpha}, \\
D T^{\alpha\dot\alpha} &=& - e_\beta{}^{\dot\alpha} R^{\alpha\beta}
- e^\alpha{}_{\dot\beta} R^{\dot\alpha\dot\beta} \nonumber
\end{eqnarray}
On-shell, i.e. on zero torsion $T^{\alpha\dot\alpha} \approx 0$, we
obtain:
\begin{equation}
D R^{\alpha(2)} \approx 0, \qquad D R^{\dot\alpha(2)} \approx 0,
\qquad e_\beta{}^{\dot\alpha} R^{\alpha\beta} + e^\alpha{}_{\dot\beta}
R^{\dot\alpha\dot\beta} \approx 0
\end{equation}
Using these curvatures the free Lagrangian can be rewritten in the
explicitly gauge invariant form:
\begin{equation}
{\cal L}_0 = \frac{i}{2\lambda^2}[ R_{\alpha(2)} R^{\alpha(2)} -
R_{\dot\alpha(2)} R^{\dot\alpha(2)} ]
\end{equation}

\subsection{Massive spin $\tz$}

The frame-like gauge invariant description uses the one-forms 
$\Phi^\alpha$, $\Phi^{\dot\alpha}$ and zero-forms $\phi^\alpha$, 
$\phi^{\dot\alpha}$. Free Lagrangian in $AdS_4$:
\begin{eqnarray}
{\cal L}_0 &=& - \Phi_\alpha e^\alpha{}_{\dot\alpha} D
\Phi^{\dot\alpha} - \phi_\alpha E^\alpha{}_{\dot\alpha} D
\phi^{\dot\alpha} \nonumber \\
 && - M \Phi_\alpha E^\alpha{}_\beta \Phi^\beta
 + c_0 \Phi_\alpha E^\alpha{}_{\dot\alpha} \phi^{\dot\alpha}
+ M E \phi_\alpha \phi^\alpha + h.c.
\end{eqnarray}
where
$$
M^2 = m^2 + \lambda^2, \qquad c_0{}^2 = 6m^2
$$
This Lagrangian is invariant under the following gauge
transformations:
\begin{equation}
\delta \Phi^\alpha = D \zeta^\alpha + M e^\alpha{}_{\dot\alpha}
\zeta^{\dot\alpha}, \qquad \delta \phi^\alpha = c_0 \zeta^\alpha
\end{equation}
Corresponding gauge invariant curvatures (two-form and one-form) can
be constructed:
\begin{eqnarray}
{\cal F}^\alpha &=& D \Phi^\alpha + M e^\alpha{}_{\dot\alpha}
\Phi^{\dot\alpha} - \frac{c_0}{3} E^\alpha{}_\beta \phi^\beta
\nonumber \\
{\cal C}^\alpha &=& D \phi^\alpha - c_0 \Phi^\alpha + M 
e^\alpha{}_{\dot\alpha} \phi^{\dot\alpha}
\end{eqnarray}
They also satisfy their differential identities:
\begin{eqnarray}
D {\cal F}^\alpha &=& - M e^\alpha{}_{\dot\alpha} 
{\cal F}^{\dot\alpha} - \frac{c_0}{3} E^\alpha{}_\beta {\cal C}^\beta
\nonumber \\ 
D {\cal C}^\alpha &=& - c_0 {\cal F}^\alpha - M 
e^\alpha{}_{\dot\alpha} {\cal C}^{\dot\alpha}
\end{eqnarray}
It is important to what follows that on-shell these curvatures can be
expressed in terms of the zero-forms \cite{KhZ19}:
\begin{equation}
\label{shell}
{\cal F}^\alpha = E_{\beta(2)} Y^{\alpha\beta(2)}, \qquad
{\cal C}^\alpha = e_{\beta\dot\alpha} Y^{\alpha\beta\dot\alpha}
\end{equation}

At last, the free Lagrangian can be rewritten as:
\begin{equation}
{\cal L}_0 = c_1 {\cal F}_\alpha {\cal F}^\alpha
+ c_2 {\cal F}_\alpha e^\alpha{}_{\dot\alpha} {\cal C}^{\dot\alpha}
+ c_3 {\cal C}_\alpha E^\alpha{}_\beta {\cal C}^\beta + h.c.
\end{equation}
where
$$
c_3 = - \frac{c_1}{3}, \qquad
2Mc_1 - c_0c_2 = \frac{1}{2}
$$
The remaining ambiguity is related with an identity
\begin{eqnarray*}
0 &=& D [ {\cal F}_\alpha {\cal C}^\alpha ]
 = D {\cal F}_\alpha {\cal C}^\alpha + {\cal F}_\alpha 
D {\cal C}^\alpha \\
 &=& - c_0 {\cal F}_\alpha {\cal F}^\alpha
- 2M {\cal F}_\alpha e^\alpha{}_{\dot\alpha} {\cal C}^{\dot\alpha}
+ \frac{c_0}{3} {\cal C}_\alpha E^\alpha{}_\beta {\cal C}^\beta
\end{eqnarray*}

\section{Deformations for spin-$\tz$}

Let us consider quadratic deformations for the spin-$\tz$ curvatures:
$$
{\cal F}^\alpha \Rightarrow \hat{\cal F}^\alpha = {\cal F}^\alpha +
\Delta {\cal F}^\alpha, \qquad {\cal C}^\alpha \Rightarrow
\hat{\cal C}^\alpha = {\cal C}^\alpha + \Delta {\cal C}^\alpha 
$$
The most general ansatz appears to be:
\begin{eqnarray}
\Delta {\cal F}^\alpha &=& a_1 \omega^\alpha{}_\beta \Phi^\beta 
+ a_2 \omega^\alpha{}_\beta e^\beta{}_{\dot\alpha} \phi^{\dot\alpha}
+ a_3 e^\alpha{}_{\dot\alpha} \omega^{\dot\alpha}{}_{\dot\beta}
\phi^{\dot\beta} \nonumber \\
 && + a_4 h^\alpha{}_{\dot\alpha} \Phi^{\dot\alpha} 
+ a_5 h^\alpha{}_{\dot\alpha} e_\beta{}^{\dot\alpha} \phi^\beta
+ a_6 e^\alpha{}_{\dot\alpha} h_\beta{}^{\dot\alpha} \phi^\beta \\
\Delta {\cal C}^\alpha &=& a_7 \omega^\alpha{}_\beta \phi^\beta
+ a_8 h^\alpha{}_{\dot\beta} \phi^{\dot\beta} \nonumber
\end{eqnarray}
Naturally, these deformed curvatures cease to be gauge invariant and
the main requirement of the formalism is that they must transform
covariantly under all gauge transformations. Let consider them in
turn.\\
{\bf Supertransformations} The appropriate corrections can be directly
read from the structure of deformations:
\begin{equation}
\delta_1 \Phi^\alpha = a_1 \omega^\alpha{}_\beta \zeta^\beta +
a_4 h^\alpha{}_{\dot\alpha} \zeta^{\dot\alpha}
\end{equation}
Now we calculate the variations for the deformed curvatures and
require:
\begin{equation}
\delta \hat{\cal F}^\alpha = a_1 R^\alpha{}_\beta \zeta^\beta + a_4
T^\alpha{}_{\dot\alpha} \zeta^{\dot\alpha}, \qquad
\delta \hat{\cal C}^\alpha = 0
\end{equation}
this gives us a number of equations on the free parameters:
$$
a_2 = a_3, \qquad a_5 = a_6, \quad
Ma_4 + c_0a_5 = a_1\lambda^2, 
$$
$$
Ma_1 + c_0a_2 = a_4, \quad a_7 = a_1, \quad a_8 = a_4
$$
{\bf Lorentz transformations} In this case the corrections have the
form:
\begin{eqnarray}
\delta \Phi^\alpha &=& - a_1 \eta^\alpha{}_\beta \Phi^\beta - a_2
\eta^\alpha{}_\beta e^\beta{}_{\dot\alpha} \phi^{\dot\alpha},
\nonumber \\
\delta \Phi^{\dot\alpha} &=& a_3 e_\alpha{}^{\dot\alpha}
\eta^\alpha{}_\beta \phi^\beta, \qquad
\delta \phi^\alpha = - a_7 \eta^\alpha{}_\beta \phi^\beta
\end{eqnarray}
For the deformed curvatures to be covariant we mast apply:
$$
a_7 = a_1, \qquad - a_8 + c_0a_2 = - Ma_7
$$
{\bf Pseudotranslations} At last, here the corrections look like:
\begin{eqnarray}
\delta \Phi^\alpha &=& - a_4 \xi^\alpha{}_{\dot\alpha}
\Phi^{\dot\alpha} - a_5 \xi^\alpha{}_{\dot\alpha} 
e_\beta{}^{\dot\alpha} \phi^\beta + a_6 e^\alpha{}_{\dot\alpha}
\xi_\beta{}^{\dot\alpha} \phi^\beta, \nonumber \\
\delta \phi^\alpha &=& - a_8 \xi^\alpha{}_{\dot\alpha}
\phi^{\dot\alpha}
\end{eqnarray}
and we obtain the last set of equations.
$$
a_8 = a_4, \qquad c_0a_5 - \lambda^2a_7 = - Ma_8, \qquad 
\lambda^2a_7 - c_0a_6 - Ma_8 = 0
$$
All the equations obtained appear to be consistent and their general
solution has two free parameters. It is not an accident because due to
the presence of zero-forms $\phi^\alpha$, $\phi^{\dot\alpha}$ there
exists a pair of possible field redefinitions, namely:
\begin{equation}
\Phi^\alpha \Rightarrow \Phi^\alpha + \kappa_1 \omega^\alpha{}_\beta
\phi^\beta + \kappa_2 h^\alpha{}_{\dot\alpha} \phi^{\dot\alpha}
\end{equation}
It is straightforward to calculate their effect on the curvatures:
\begin{eqnarray*}
\Delta {\cal F}^\alpha &=& \kappa_1
D \omega^\alpha{}_\beta \phi^\beta - \kappa_1 \omega^\alpha{}_\beta
D \phi^\beta + M\kappa_1 e^\alpha{}_{\dot\alpha}
\omega^{\dot\alpha}{}_{\dot\beta} \phi^{\dot\beta} \\
 &=& - \kappa_1 R^{\alpha\beta} \phi_\beta - \kappa_1
\omega^\alpha{}_\beta {\cal C}^\beta \\
 && - \kappa_1c_0 \omega^\alpha{}_\beta \Phi^\beta 
 + M\kappa_1 ( \omega^\alpha{}_\beta e^\beta{}_{\dot\alpha} +
e^\alpha{}_{\dot\beta} \omega^{\dot\beta}{}_{\dot\alpha})
\phi^{\dot\alpha} \\
 && - \kappa_1\lambda^2 (h^\alpha{}_{\dot\alpha}
e_\beta{}^{\dot\alpha} + e^\alpha{}_{\dot\alpha} 
h_\beta{}^{\dot\alpha}) \phi^\beta \\
\Delta {\cal C}^\alpha &=& - \kappa_1c_0
\omega^\alpha{}_\beta \phi^\beta
\end{eqnarray*}
\begin{eqnarray*}
\Delta {\cal F}^\alpha &=& \kappa_2 
D h^\alpha{}_{\dot\alpha} \phi^{\dot\alpha} - \kappa_2
h^\alpha{}_{\dot\alpha} D \phi^{\dot\alpha} + M\kappa_2
e^\alpha{}_{\dot\alpha} h_\beta{}^{\dot\alpha} \phi^\beta \\
 &=& \kappa_2 T^\alpha{}_{\dot\alpha} \phi^{\dot\alpha} 
- \kappa_2 h^\alpha{}_{\dot\alpha} {\cal C}^{\dot\alpha} \\
 && - \kappa_2 (\omega^\alpha{}_\beta e^\beta{}_{\dot\alpha}
+ e^\alpha{}_{\dot\beta} \omega^{\dot\beta}{}_{\dot\alpha})
\phi^{\dot\alpha} - \kappa_2c_0 h^\alpha{}_{\dot\alpha}
\Phi^{\dot\alpha} \\
 &&  + M\kappa_2 (h^\alpha{}_{\dot\alpha}
e_\beta{}^{\dot\alpha} + e^\alpha{}_{\dot\alpha}
h_\beta{}^{\dot\alpha}) \phi^\beta \\
\Delta {\cal C}^\alpha &=& - \kappa_2c_0 
h^\alpha{}_{\dot\alpha} \phi^{\dot\alpha}
\end{eqnarray*}
Thus these redefinitions shift the deformation parameters:
\begin{eqnarray*}
a_{1,7} &\Rightarrow& a_{1,7} - \kappa_1c_0 \\
a_{2,3} &\Rightarrow& a_{2,3} + M\kappa_1 - \kappa_2, \\
a_{4,8} &\Rightarrow& a_{4,8} - \kappa_2c_0 \\
a_{5,6} &\Rightarrow& a_{5,6} - \kappa_1\lambda^2 + M\kappa_2
\end{eqnarray*}
and also generate some abelian corrections. All the equations on the
parameters $a_{1-8}$ given above are invariant under these shifts and
it serves as a non-trivial independent check for our calculations.
Moreover, using these redefinitions all parameters $a_{1-8}$ can be
set to zero leaving us with the abelian deformations only in agreement
with the general analysis in \cite{BDGT18}.

\section{Deformations for spin-2}

Now let us turn to the quadratic deformations for massless spin-2. The
most general ansatz looks like:
\begin{eqnarray}
\frac{1}{i}\Delta R^{\alpha(2)} &=& b_1 \Phi^\alpha \Phi^\alpha 
+ b_2 e^\alpha{}_{\dot\alpha} \Phi^\alpha \phi^{\dot\alpha}
+ b_3 e^\alpha{}_{\dot\alpha} \Phi^{\dot\alpha} \phi^\alpha \nonumber
\\
 && + b_4 E^{\alpha(2)} \phi^\beta \phi_\beta
+ b_5 E^{\alpha(2)} \phi^{\dot\beta} \phi_{\dot\beta} \nonumber \\
\frac{1}{i}\Delta T^{\alpha\dot\alpha} &=& b_6 \Phi^\alpha
\Phi^{\dot\alpha} + b_7 (e^\alpha{}_{\dot\beta} \Phi^{\dot\alpha}
\phi^{\dot\beta} + e_\beta{}^{\dot\alpha} \Phi^\alpha \phi^\beta) \\
 && + b_8 (e^\alpha{}_{\dot\beta} \Phi^{\dot\beta} \phi^{\dot\alpha} 
+ e_\beta{}^{\dot\alpha} \Phi^\beta \phi^\alpha) \nonumber \\
 && + b_9 (E^\alpha{}_\beta \phi^\beta \phi^{\dot\alpha}
- E^{\dot\alpha}{}_{\dot\beta} \phi^\alpha \phi^{\dot\beta}) \nonumber
\end{eqnarray}
Note that one more possible term $e^{\alpha\dot\alpha} (\Phi^\beta
\phi_\beta + \Phi^{\dot\beta} \phi_{\dot\beta})$ is equivalent to the
combination of the terms with parameters $b_{7,8}$.  Let us consider
the variations for the deformed curvatures and require that they
transform covariantly. \\
{\bf $\zeta^\alpha$-transformations} Here the corrections to
supertransformations have the form:
\begin{eqnarray}
\delta_1 \omega^{\alpha(2)} &=& 2b_1 \Phi^\alpha \zeta^\alpha +
b_2 e^\alpha{}_{\dot\alpha} \zeta^\alpha \phi^{\dot\alpha}, \nonumber
\\
\delta_1 h^{\alpha\dot\alpha} &=& b_6 \Phi^{\dot\alpha} \zeta^\alpha
- b_7 e_\beta{}^{\dot\alpha} \phi^\beta \zeta^\alpha
- b_8 e_\beta{}^{\dot\alpha} \phi^\alpha \zeta^\beta
\end{eqnarray}
and the covariance of the deformed curvatures gives a number of
equations on the parameters:
\begin{eqnarray*}
&& c_0b_3 + \lambda^2 b_6 = 2Mb_1 - c_0b_2 \\
&& - 2Mb_3 - c_0b_4 - 2\lambda^2b_8 = 0 \\
&& c_0b_4 - 2\lambda^2 b_7 = - \frac{2b_1c_0}{3} + 2Mb_2 \\
&& 2b_1 - Mb_6 + c_0b_7 + c_0b_8 = 0 \\
&& 2b_1 - Mb_6 + c_0b_7 + c_0b_8 = 0 \\
&& - Mb_7 - Mb_8 - c_0b_9 + b_2 - 3b_3 = 0   \\
&& - Mb_7 - Mb_8 + c_0b_9 - 3b_2 + b_3 = - \frac{c_0b_6}{3} 
\end{eqnarray*}
{\bf $\zeta^{\dot\alpha}$-transformations} In this case we have:
\begin{eqnarray}
\delta_1 \omega^{\alpha(2)} &=& - b_3 e^\alpha{}_{\dot\alpha}
\phi^\alpha \zeta^{\dot\alpha}, \nonumber \\
\delta_1 h^{\alpha\dot\alpha} &=& b_6 \Phi^\alpha \zeta^{\dot\alpha} 
- b_7 e^\alpha{}_{\dot\beta} \phi^{\dot\beta} \zeta^{\dot\alpha}
- b_8 e^\alpha{}_{\dot\beta} \phi^{\dot\alpha} \zeta^{\dot\beta}
\end{eqnarray}
and obtain a couple of additional equations:
\begin{eqnarray*}
&& - 2Mb_1 + c_0b_2 + \lambda^2b_6 = - c_0b_3 \\
&& 2Mb_2 + 2c_0b_5 - 2\lambda^2b_7 + 2\lambda^2b_8 = 2Mb_3
\end{eqnarray*}
As in the previous case, these equations appear to be consistent and
their general solution has four arbitrary parameters. Here this is
also
connected with the existence of four possible field redefinitions:
\begin{eqnarray}
\omega^{\alpha(2)} &\Rightarrow& \omega^{\alpha(2)} + i\rho_1 
\Phi^\alpha \phi^\alpha + i\rho_2 e^\alpha{}_{\dot\alpha}
\phi^\alpha \phi^{\dot\alpha} \nonumber \\
\omega^{\dot\alpha(2)} &\Rightarrow& \omega^{\dot\alpha(2)} + i\rho_1
\Phi^{\dot\alpha} \phi^{\dot\alpha} + i\rho_2 e_\alpha{}^{\dot\alpha}
\phi^{\dot\alpha} \phi^\alpha \\
h^{\alpha\dot\alpha} &\Rightarrow& h^{\alpha\dot\alpha} 
+ i\rho_3 (\Phi^\alpha \phi^{\dot\alpha} + \Phi^{\dot\alpha}
\phi^\alpha) + i\rho_4 e^{\alpha\dot\alpha} (\phi^\beta \phi_\beta +
\phi^{\dot\beta} \phi_{\dot\beta}) \nonumber
\end{eqnarray}
Their effects on the curvatures:
\begin{eqnarray*}
\frac{1}{i\rho_1} \Delta R^{\alpha(2)} &=& D \Phi^\alpha \phi^\alpha 
- \Phi^\alpha D \phi^\alpha \\
 &=&  {\cal F}^\alpha \phi^\alpha
-  \Phi^\alpha {\cal C}^\alpha - c_0 \Phi^\alpha
\Phi^\alpha \\
 && - M  e^\alpha{}_{\dot\alpha} [\Phi^\alpha \phi^{\dot\alpha}
+ \Phi^{\dot\alpha} \phi^\alpha] + \frac{c_0}{3}
E^{\alpha(2)} \phi^\beta \phi_\beta   \\
\frac{1}{i} \Delta T^{\alpha\dot\alpha} &=& 
e_\beta{}^{\dot\alpha} ( \Phi^\alpha \phi^\beta + \Phi^\beta
\phi^\alpha) + \rho_1 e^\alpha{}_{\dot\beta} (\Phi^{\dot\alpha}
\phi^{\dot\beta} + \Phi^{\dot\beta} \phi^{\dot\alpha})
\end{eqnarray*}
\begin{eqnarray*}
\frac{1}{i\rho_2} \Delta R^{\alpha(2)} &=& - e^\alpha{}_{\dot\alpha} D
\phi^\alpha \phi^{\dot\alpha} - e^\alpha{}_{\dot\alpha} \phi^\alpha
D \phi^{\dot\alpha} \\
 &=& - e^\alpha{}_{\dot\alpha} {\cal C}^\alpha \phi^{\dot\alpha} -
e^\alpha{}_{\dot\alpha} \phi^\alpha {\cal C}^{\dot\alpha} 
  - c_0 e^\alpha{}_{\dot\alpha} (\Phi^\alpha \phi^{\dot\alpha}
- \Phi^{\dot\alpha} \phi^\alpha) \\
 &&  - 2M E^{\alpha(2)} \phi^\beta
\phi_\beta + 2M E^{\alpha(2)} \phi^{\dot\alpha} \phi_{\dot\alpha}  \\
\frac{1}{i\rho_2} \Delta T^{\alpha\dot\alpha} &=& 
 - 4E^\alpha{}_\beta \phi^\beta \phi^{\dot\alpha}
+ 4E^{\dot\alpha}{}_{\dot\beta} \phi^\alpha \phi^{\dot\beta} 
\end{eqnarray*}
\begin{eqnarray*}
\frac{1}{i\rho_3} \Delta R^{\alpha(2)} &=& \lambda^2 
e^\alpha{}_{\dot\alpha} (\Phi^\alpha \phi^{\dot\alpha} +
\Phi^{\dot\alpha} \phi^\alpha) \\
\frac{1}{i\rho_3} \Delta T^{\alpha\dot\alpha} &=& D \Phi^\alpha
\phi^{\dot\alpha} - \Phi^\alpha D \phi^{\dot\alpha} +
D \Phi^{\dot\alpha} \phi^\alpha - \Phi^{\dot\alpha} D \phi^\alpha \\
 &=& {\cal F}^\alpha \phi^{\dot\alpha} - \Phi^\alpha 
{\cal C}^{\dot\alpha} + {\cal F}^{\dot\alpha} \phi^\alpha -
\Phi^{\dot\alpha} {\cal C}^\alpha - 2c_0 \Phi^\alpha \Phi^{\dot\alpha}
\\
 &&  - M e_\beta{}^{\dot\alpha}
(\Phi^\alpha \phi^\beta + \Phi^\beta \phi^\alpha) 
- M e^\alpha{}_{\dot\beta} (\Phi^{\dot\alpha} \phi^{\dot\beta} +
\Phi^{\dot\beta} \phi^{\dot\alpha}) \\
 && + \frac{c_0}{3} E^\alpha{}_\beta
\phi^\beta \phi^{\dot\alpha} - \frac{c_0}{3}
E^{\dot\alpha}{}_{\dot\beta} \phi^\alpha \phi^{\dot\beta}
\end{eqnarray*}
\begin{eqnarray*}
\frac{1}{i\rho_4} \Delta R^{\alpha(2)} &=& 4\lambda^2 
E^{\alpha(2)} (\phi^\beta \phi_\beta + \phi^{\dot\beta}
\phi_{\dot\beta}) \\
\frac{1}{i\rho_4} \Delta T^{\alpha\dot\alpha} &=& - 2
e^{\alpha\dot\alpha} (D \phi^\beta \phi_\beta + D \phi^{\dot\beta}
\phi_{\dot\beta}) \\
  &=& - 2 e^{\alpha\dot\alpha} ( {\cal C}^\beta \phi_\beta +
{\cal C}^{\dot\beta} \phi_{\dot\beta}) \\
 && - 2c_0 e^{\alpha\dot\alpha} (\Phi^\beta \phi_\beta +
\Phi^{\dot\beta} \phi_{\dot\beta}) 
\end{eqnarray*}
Thus they also produce the shifts of the deformation parameters:
\begin{eqnarray*}
b_1 &\Rightarrow& b_1 - \rho_1c_0  \\
b_2 &\Rightarrow& b_2 - \rho_1 M - \rho_2c_0 + \rho_3\lambda^2 \\
b_3 &\Rightarrow& b_3 - \rho_1 M + \rho_2c_0 + \rho_3\lambda^2 \\
b_4 &\Rightarrow& b_4 + \frac{\rho_1c_0}{3} - 2\rho_2M +
4\rho_4\lambda^2 \\
b_5 &\Rightarrow& b_5 + 2\rho_2M + 4\rho_4\lambda^2 \\
b_6 &\Rightarrow& b_6 - 2\rho_3c_0  \\
b_7 &\Rightarrow& b_7 + \rho_1 - \rho_3M + 2\rho_4c_0 \\
b_8 &\Rightarrow& b_8 + \rho_1 - \rho_3M - 2\rho_4c_0 \\
b_9 &\Rightarrow& b_9 - 4\rho_2 + \frac{\rho_3c_0}{3}
\end{eqnarray*}
and generate a number of abelian deformations. We have explicitly
checked that all equations are invariant under these shifts. Moreover,
using these redefinitions one can set all deformation parameters
$b_{1-9}$ to zero, once again in agreement with the general analysis
in
\cite{BDGT18}.

\section{Abelian vertices}

As we have seen in the previous two subsections, using all field
redefinitions we can reformulate the cubic interactions as a
combination of the abelian and/or trivially gauge invariant terms.
Note first of all, that there exists just one trivially gauge
invariant vertex that does not vanish on-shell:
\begin{equation}
{\cal L}_t \sim R_{\alpha\beta} {\cal C}^\alpha {\cal C}^\beta + h.c.
\end{equation}
Formally, it contains terms with up to four derivatives, but up to the
total derivative it is equivalent to the particular combination of the
abelian ones, as can be seen from
\begin{eqnarray*}
{\cal L} &=& [ D \omega_{\alpha\beta} - \lambda^2 
(e_\alpha{}^{\dot\alpha} h_{\beta\dot\alpha} + e_\beta{}^{\dot\alpha}
h_{\alpha\dot\alpha}) ] {\cal C}^\alpha {\cal C}^\beta + h.c. \\
 &=& 2c_0 {\cal F}_\alpha {\cal C}_\beta \omega^{\alpha\beta} 
 - 2M {\cal C}_\alpha {\cal C}_{\dot\alpha} e_\beta{}^{\dot\alpha}
\omega^{\alpha\beta} + 2\lambda^2  {\cal C}_\alpha {\cal C}_\beta
e^\alpha{}_{\dot\alpha} h^{\beta\dot\alpha} + h.c.
\end{eqnarray*}
Recall that we call abelian vertices constructed out of two gauge
invariant curvatures and one explicit field. Thus in our case we have
two types of abelian vertices, namely those with the massive 
spin-$\tz$ components $\Phi^\alpha$ or $\phi^\alpha$ and those with
$h^{\alpha\dot\alpha}$ or $\omega^{\alpha(2)}$. As for the first ones,
it is easy to show that any gauge invariant combinations of them is
on-shell equivalent to the trivially gauge invariant one given above,
once again in agreement with \cite{BDGT18}. But for the vertices with
the massless spin-2 the situation appears to be different. The most
general ansatz for such vertices is:
\begin{equation}
{\cal L}_a = d_1 {\cal F}_\alpha {\cal C}_\beta \omega^{\alpha\beta}
+ d_2 {\cal C}_\alpha {\cal C}_{\dot\alpha} e_\beta{}^{\dot\alpha}
\omega^{\alpha\beta} + d_3 {\cal F}_\alpha {\cal C}_{\dot\alpha}
h^{\alpha\dot\alpha} + d_4 {\cal C}_\alpha {\cal C}_\beta
e^\alpha{}_{\dot\alpha} h^{\beta\dot\alpha} + h.c.
\end{equation}
If we require that these terms be gauge invariant, then we find that
all $\eta^{\alpha(2)}$-variations vanish on-shell (\ref{shell}), while
for the $\xi^{\alpha\dot\alpha}$-transformations we obtain
\begin{equation}
\delta_\xi {\cal L}_a \approx [\lambda^2d_1 - Md_3 - c_0d_4] 
{\cal F}_\alpha {\cal C}_\beta e^\beta{}_{\dot\alpha}
\xi^{\alpha\dot\alpha}
\end{equation}
Thus we have only one equation on four parameters:
$$
\lambda^2 d_1 - Md_3 - c_0d_4 = 0
$$
and this leaves with the three possible solutions. One of them as it
was shown above is equivalent to the trivially gauge invariant one,
but two other ones are independent and have to be considered
separately. The most simple way to determine which combination
reproduces the minimal vertex is to consider the so-called unitary
gauge ($\phi^\alpha=0$)\footnote{The reason is that all field
redefinitions are necessarily contain zero-forms $\phi^\alpha$ and do
not change this part of the vertex.}. Using explicit expressions for
the curvatures we obtain:
\begin{eqnarray}
{\cal L}_a &=& -c_0d_1 (D \Phi_\alpha - M e_\alpha{}^{\dot\alpha}
\Phi_{\dot\alpha}) \Phi_\beta \omega^{\alpha\beta}
 + c_0{}^2d_2 \Phi_\alpha \Phi_{\dot\alpha} e_\beta{}^{\dot\alpha}
\omega^{\alpha\beta} \nonumber \\
 && - c_0d_3 (D \Phi_\alpha - M e_\alpha{}^{\dot\alpha}
\Phi_{\dot\alpha}) \Phi_{\dot\beta} h^{\alpha\dot\beta}
+ c_0{}^2d_4 \Phi_\alpha \Phi_\beta e^\alpha{}_{\dot\alpha}
h^{\beta\dot\alpha}
\end{eqnarray}
It appears that the minimal vertex having no more than one derivative
corresponds to the choice:
$$
d_1 = 0, \qquad d_3 = c_0d_2, \qquad d_4 = - Md_2
$$
As a result we obtain
\begin{equation}
\frac{1}{c_0{}^2} {\cal L}_a = d_2 [ \Phi_\alpha \Phi_{\dot\alpha}
e_\beta{}^{\dot\alpha} \omega^{\alpha\beta} - D\Phi_\alpha
\Phi_{\dot\alpha} h^{\alpha\dot\alpha} - 2M \Phi_\alpha \Phi_\beta
e^\alpha{}_{\dot\alpha} h^{\beta\dot\alpha} + h.c.]
\end{equation}
and this indeed agrees with the minimal substitution rule for the
corresponding part of the free Lagrangian. We have explicitly checked
that all the terms with two derivatives (beyond the unitary gauge)
vanish on-shell.

\section{Comeback}

Thus we managed to reproduce the minimal gravitational vertex for
massive spin-$\tz$ as a purely abelian one. Abelian vertices are easy
to construct and simple to deal with, but such a form drastically
differs from what we used to working with the massless fields where
the most important part (and in $d=4$ very often the only one) is
non-abelian one. So it seems instructive to bring the result in the
form close to that of massless field, in-particular to see can the
massive theory be interpreted as the spontaneously broken massless one
as the presence of the Stueckelberg fields suggests.

As the first step we use the field redefinition to restore the
non-abelian part of the curvature deformations. We chose
\begin{eqnarray}
\Delta {\cal F}^\alpha &=& a_1 [ \omega^\alpha{}_\beta \Phi^\beta +
\kappa_0 \omega^\alpha{}_\beta e^\beta{}_{\dot\alpha}
\phi^{\dot\alpha} + \kappa_0 e^\alpha{}_{\dot\alpha}
\omega^{\dot\alpha}{}_{\dot\beta} \phi^{\dot\beta}] \nonumber \\
 && + \lambda a_1 [ h^\alpha{}_{\dot\alpha} \Phi^{\dot\alpha} +
\kappa_0 h^\alpha{}_{\dot\alpha} e_\beta{}^{\dot\alpha} 
\phi^\beta + \kappa_0 e^\alpha{}_{\dot\alpha} h_\beta{}^{\dot\alpha}
\phi^\beta ] \\
\Delta {\cal C}^\alpha &=& a_1 \omega^\alpha{}_\beta \phi^\beta +
\lambda a_1 h^\alpha{}_{\dot\alpha} \phi^{\dot\alpha} \nonumber
\end{eqnarray}
for the spin-$\tz$ curvatures, where
$$
\kappa_0 = - \frac{M-\lambda}{c_0}
$$
while for the spin-2 ones:
\begin{eqnarray}
\Delta R^{\alpha(2)} &=& b_1 [\Phi^\alpha \Phi^\alpha - 2\kappa_0
e^\alpha{}_{\dot\alpha} \Phi^\alpha \phi^{\dot\alpha} - 2\kappa_0{}^2
E^{\alpha(2)} \Phi^{\dot\beta} \phi_{\dot\beta} \nonumber \\
\Delta T^{\alpha\dot\alpha} &=& b_6 [ \Phi^\alpha \Phi^{\dot\alpha} -
\kappa_0 (e^\alpha{}_{\dot\beta} \Phi^{\dot\alpha} \phi^{\dot\beta} +
e_\beta{}^{\dot\alpha} \Phi^\alpha \phi^\beta) \\
 && - \kappa_0{}^2 (E^\alpha{}_\beta \phi^\beta \phi^{\dot\alpha} -
E^{\dot\alpha}{}_{\dot\beta} \phi^\alpha \phi^{\dot\beta})] \nonumber
\end{eqnarray}
$$
2b_1 = \lambda b_6 = - 2c_1a_1 \lambda^2
$$
The reason for these particular choice will become clear in a moment.

As our second step let us introduce new variable:
$$
\tilde\Phi^\alpha = \Phi^\alpha + \kappa_0 e^\alpha{}_{\dot\alpha}
\phi^{\dot\alpha}
$$
so that it transforms exactly as the massless spin-$\tz$ field. 
$$
\delta \tilde\Phi^\alpha = D \zeta^\alpha + \lambda 
e^\alpha{}_{\dot\alpha} \zeta^{\dot\alpha}
$$
In turn, for the gauge invariant curvatures we obtain:
\begin{eqnarray}
\tilde{\cal F}^\alpha &=& {\cal F}^\alpha - \kappa_0 
e^\alpha{}_{\dot\alpha} {\cal C}^{\dot\alpha} =
D \tilde\Phi^\alpha + \lambda e^\alpha{}_{\dot\alpha}
\tilde\Phi^{\dot\alpha} \nonumber \\
{\cal C}^\alpha &=& D \phi^\alpha - c_0 \tilde\Phi^\alpha + \lambda
e^\alpha{}_{\dot\alpha} \phi^{\dot\alpha}
\end{eqnarray}
Now we at last fix our choice for the form of the free Lagrangian
written in terms of curvatures. We choose
$$
c_2 = - 2c_1\kappa_0 \quad \Rightarrow \quad c_1 = \frac{1}{4\lambda}
$$
then the free Lagrangian takes the form:
\begin{equation}
{\cal L}_0 = \frac{1}{4\lambda} \tilde{\cal F}_\alpha
\tilde{\cal F}^\alpha - \frac{1}{6(M+\lambda)} {\cal C}_\alpha
E^\alpha{}_\beta {\cal C}^\beta + h.c.
\end{equation}
so that the first term coincides with the usual Lagrangian for
massless spin-$\tz$ field in $AdS_4$, while the second one contains
mass terms dressed with the Stueckelberg fields.

In these new variables the quadratic deformations take the form
\begin{eqnarray}
\Delta \tilde{\cal F}^\alpha &=& a_1 [ \omega^\alpha{}_\beta
\tilde\Phi^\beta + \lambda h^\alpha{}_{\dot\alpha}
\tilde\Phi^{\dot\alpha}] \nonumber \\
\Delta {\cal C}^\alpha &=& a_1 [ \omega^\alpha{}_\beta \phi^\beta +
\lambda h^\alpha{}_{\dot\alpha} \phi^{\dot\alpha} ]
\end{eqnarray}
\begin{equation}
\Delta R^{\alpha(2)} = - \frac{a_1\lambda}{4} \tilde\Phi^\alpha
\tilde\Phi^\alpha, \qquad
\Delta T^{\alpha\dot\alpha} = - \frac{a_1}{2} \tilde\Phi^\alpha
\tilde\Phi^{\dot\alpha} 
\end{equation}
while the interacting Lagrangian looks like
\begin{eqnarray}
{\cal L} &=& \frac{i}{2\lambda^2} \hat{R}_{\alpha(2)} 
\hat{R}^{\alpha(2)} + \frac{1}{4\lambda} \hat{\tilde{\cal F}}_\alpha
\hat{\tilde{\cal F}^\alpha} \nonumber \\
 && - \frac{1}{6(M+\lambda)} 
\hat{\cal C}_\alpha E^\alpha{}_\beta \hat{\cal C}^\beta 
+ \frac{a_1}{12(M+\lambda)} {\cal C}_\alpha {\cal C}_\beta
e^\alpha{}_{\dot\alpha} h^{\beta\dot\alpha} + h.c.
\end{eqnarray}
where the first two terms coincide with corresponding results for the
massless supergravity. Note that we still have a combination of the
non-abelian deformation and one abelian vertex.

\section{Conclusion}

In this work we applied the so-called Fradkin-Vasiliev formalism to
the system of massless spin-2 and massive spin-$\tz$ fields.
Initially, this formalism was developed for the construction of
interactions for massless higher spins, but using the frame-like gauge
invariant description for massive ones, it can be straightforwardly
extended to any system of massless and/or massive fields. In massless
case, the interaction vertices appears as a result of most general
quadratic deformation for all gauge invariant curvatures in
combination with all possible abelian (or Chern-Simons like) terms.
But in the massive case \cite{BDGT18} due to the presence of
Stueckelberg fields there always exist enough field redefinition to
bring the vertex into purely abelian form. We have shown here that
this statement remains valid in the presence of massless spin-2 (with
its unbroken gauge symmetries) as well. At the same time, our
investigation of the abelian vertices show that there exist two such
vertices which are not equivalent on-shell to any trivially gauge
invariant ones contrary to the case where only massive fields are
present. Moreover, it is one of these vertices that reproduce standard
minimal gravitational cubic vertex for massive spin-$\tz$. We leave to
the future work the extension of such investigations to other higher
spins fields, first of all to massive (bi)gravity and supergravity.

\section*{Acknowledgements}
M.Kh. is grateful to Foundation for the Advancement of Theoretical
Physics and Mathematics "BASIS" for their support of the work.

\end{document}